\documentclass[aps,pra,twocolumn,amssymb]{revtex4}
\usepackage{epsfig}
\begin{document}
\title{KLM quantum computation with bosonic atoms}

\author{Sandu Popescu$^{a,b}$}
\affiliation{$^c$H. H. Wills Physics Laboratory, University of Bristol, Tyndall Avenue, Bristol BS8 1TL}
\affiliation{$^b$Hewlett-Packard Laboratories, Stoke Gifford, Bristol BS12 6QZ, UK}
\date{\today}

\begin{abstract}
A  Knill-Laflamme-Milburn (KLM) type quantum computation with bosonic neutral atoms or bosonic ions is
suggested. Crucially, as opposite to other quantum computation schemes involving atoms (ions), no controlled
interactions between atoms (ions) involving their internal levels are required. Versus photonic KLM computation,
this scheme has the advantage that  single atom (ion) sources are more natural than single photon sources, and
single atom (ion) detectors are far more efficient than single photon ones.

\end{abstract}

\pacs{PACS numbers: 03.67.-a}

\maketitle

\newcommand{\beq}{\begin{equation}}
\newcommand{\eeq}{\end{equation}}
\newcommand{\ra}{\rangle}
\newcommand{\la}{\langle}

The purpose of this short note is to point out that one can perform Knill, Laflamme and Millburn (KLM) type
quantum computation \cite{klm} not only with photons but also with bosonic neutral atoms or with bosonic ions.
Such a scheme has a number of practical advantages both over optical KLM computation as well as over traditional
quantum computation using neutral atoms/ions.

At present there are many proposals for performing quantum computation with both neutral atoms as well as with
ions. All these proposals require carefully controlled interactions between the neutral atoms (ions) and involve
manipulating their internal states. In contrast, the KLM type computation presented below does not involve any
interaction between the internal degree of freedoms of the atoms(ions). In fact all interactions are actually
avoided! The computation involves only manipulating their center of mass.

Conceptually, KLM computation with neutral bosonic atoms and bosonic ions are identical, so, to simplify
terminology,  in what follows I will use the word ``atoms" to represent both neutral bosonic atoms and bosonic
ions.   It is however worthwhile mentioning right from the beginning that due to the much stronger interaction
that ions have with each other and with the environment KLM computation with ions is probably impractical.

One of the major stumbling blocks in KLM quantum computation is the need for deterministic single photon sources
which act as the input state for the computation. Laser pulses, that are easy to produce on demand, are not
single photon states - they are coherent states, i.e. they are in a superposition of different photon numbers.
One way to prepare single photon states is by parametric down-conversion, a process in which one ultra-violet
photon impinging on a suitable crystal has a probability of being converted into a pair of optical photons.
Detecting one of the photons in a pair guarantees that one optical photon (its partner)  is present. However,
this process is probabilistic, and gives only a-posteriory information that a state containing one optical
photon has been produced. True single photon deterministic sources have already been constructed \cite{sources}
but are at the limit of present day technology, and generating in a synchonised way a large number of single
photon states seems quite remote. On the other hand, by their very nature, atoms are always found in single (or
well-determined number) states, so preparing the input state should be much easier.

A second major practical problem in photon KLM computation is the need for single-photon detectors which perform
the measurement of the final and certain intermediate states. Such detectors are crucial for driving the
computation (by feed forward). But single photon detectors are notoriously inefficient; they are subjected both
to losses (when a photon reaches the detector but the detector doesn't click) as well as to ``dark counts" (when
no photon is present but the detector nevertheless clicks). On the other hand, detecting atoms with high
efficiency is rather easy.

A KLM computation with bosonic atoms should take place in a very similar way to a computation with photons.
Whenever in a photon computation we inject a photon, we now shoot an atom. Optical mirrors and beam-splitters
are replaced with their equivalents for atoms, and can be realized by appropriate arrangements of electric
fields or laser beams.

As a technicality, to avoid instabilities affecting the trajectory of the atoms  as they travel through the
complex multi-particle interferometer that is a KLM computer, one can imagine that individual atoms are captured
in the ground-state of individual potential wells, and then the potential wells are moved and drag the captured
atoms with them. Of course, there will be instances when we do not know where an atom  is since the state can be
a superposition of the atom  being in many different locations. This happens, for example, after an atom
impinges onto a beam-splitter. To insure the guiding effect we must use a moving potential well associated to
each incoming/outgoing mode (see fig. 1). The atom  will then end up in a superposition of being in different
potential wells.

There are already various practical ways of guiding both neutral atoms as well as ions and the technology is
improving very fast.

\begin{figure}[h]
\epsfig{file=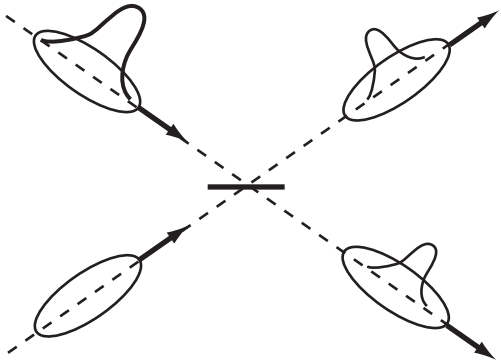} \caption{The diagram shows an atomic wavepacket impinging onto a beamsplitter guided by
moving potential wells. Note that there is one potential well per mode.}
\end{figure}

At first sight it seems that KLM computation with atoms cannot work because KLM computation implies ``linear"
interactions. More precisely, photons do not scatter when they meet, and at each element (mirror, beamsplitter,
etc.) each photon behaves as if other photons are not present. For example if say, a piece of apparatus makes a
single photon state evolve according to \beq a^{\dagger}|0>\rightarrow b^{\dagger}|0>,\eeq then an $n$ photon
state evolves according to
 \beq (a^{\dagger})^n|0>\rightarrow (b^{\dagger})^n|0>\eeq
On the other hand, unlike photons, atoms interact with each other and scatter each other. Hence one might expect
that the evolution of atoms  in a KLM computer will, in general, be completely different from that of photons. I
will argue however that this need not be a problem and that one can engineer a regime in which an atom KLM
interferometer works almost identically to a photon KLM interferometer.

The main effect in KLM computation is the interference that occurs when two photons impinge simultaneously on a
beamsplitter, one photon from each side. Let $a^{\dagger}_1$ and $a^{\dagger}_2$ denote the incoming modes and
$b^{\dagger}_1$ and $b^{\dagger}_2$ denote the outgoing modes. Suppose first that a single photon impinges on
the beamsplitter (considered to be 50\%-50\%), in mode $a^{\dagger}_1$ . The standard evolution is \beq
a^{\dagger}_1|0>\rightarrow {1\over{\sqrt 2}}(b^{\dagger}_1+b^{\dagger}_2)|0>.\label{mode1}\eeq On the other
hand, when a single photon in mode $a^{\dagger}_2$ impinges on the beamsplitter, the evolution is \beq
a^{\dagger}_2|0>\rightarrow {1\over{\sqrt 2}}(b^{\dagger}_1-b^{\dagger}_2)|0>.\label{mode2}\eeq Now, when two
photons impinge together on the beamsplitter, the evolution is \beq a^{\dagger}_2 a^{\dagger}_1|0>\rightarrow
{1\over{\sqrt 2}}(b^{\dagger}_1-b^{\dagger}_2)|{1\over{\sqrt 2}}(b^{\dagger}_1+b^{\dagger}_2)|0>\nonumber\eeq
\beq={1\over 2}\bigl((b^{\dagger}_1)^2-(b^{\dagger}_2)^2\bigr)|0>.\label{mandeldip}\eeq Hence, the photons
emerging from the beamsplitter are correlated, in a superposition of both being in mode $b^{\dagger}_1$ or both
in mode $b^{\dagger}_2$, that is, both leave the beamsplitter in the same direction. (In quantum optics this
effect is known as the Hong-Ou-Mandel dip\cite{mandel_dip}).

Suppose however that we are dealing with atoms instead of photons. When a single atom impinges on an (atomic)
beamsplitter, the evolution is similar to that described in equations (\ref{mode1}) and (\ref{mode2}). On the
other hand, when two atoms reach the beamsplitter simultaneously they will colide and scatter so that instead of
the interference effect described in (\ref{mandeldip}), we expect the evolution to be described by \beq
a^{\dagger}_2 a^{\dagger}_1|0>\rightarrow |scattering ~state>.\label{atomsmandeldip}\eeq The atoms will not
emerge both in the same mode but may end up one on each side of the beamsplitter, or scattered in other
directions altogether. Since the Hong-Ou-Mandel dip interference is the basic effect that makes KLM computation
work, it seems that there is no way to implement it with atoms.

The situation however is not so bad. To see this we need to look more carefully at the incoming wave-packets,
and analyze in detail how the scattering occurs. Suppose that the incoming modes $a^{\dagger}_1$ and
$a^{\dagger}_2$ represent wave-packets that are much longer than the scattering cross-section of the atoms. Let
us decompose these long wave-packets into shorter wave-packets, each short wave-packet having the dimension of
the scattering length. Let these short wave-packets be described by $a^{\dagger}_{1,t_i}$ and
$a^{\dagger}_{2,t_i}$, and the corresponding outgoing modes be described by $b^{\dagger}_{1,t_i}$ and
$b^{\dagger}_{2,t_i}$, . The index $t_i$ may be thought of indicating the time the wave-packet reaches the
beamsplitter. Then

 \beq a^{\dagger}_{1}= {1\over{\sqrt n}}\sum_{i=1}^n a^{\dagger}_{1,t_i}\nonumber\eeq

\beq a^{\dagger}_{2}= {1\over{\sqrt n}}\sum_{i=1}^n a^{\dagger}_{2,t_i,}\eeq where $n={L\over l}$ represents the
number of short wave-packets (length $l$) that make the long wave-packet (length $L$).

Suppose now that instead of the long wave-packets $a^{\dagger}_{1}$ and $a^{\dagger}_{2}$ we send towards the
beamsplitter only two short wavepackets, $a^{\dagger}_{1,t_i}$ and  $a^{\dagger}_{2,t_j}$. If $i=j$, the two
wave-packets arrive at the beamsplitter simultaneously and the atoms scatter

 \beq a^{\dagger}_{2,t_i}a^{\dagger}_{2,t_i}|0>\rightarrow |scattered~state_i>.\eeq
On the other hand, when $i\neq j$ the atoms arive at the beamsplitter at different times. Since they arive such
that the distance between them is larger than the scattering lengt, they do not disturb each other and they
evolve independently, in he same way as photons would do, namely \beq
a^{\dagger}_{2,t_j}a^{\dagger}_{2,t_i}|0>\rightarrow {1\over{\sqrt 2}}(b^{\dagger}_{1,
t_j}-b^{\dagger}_{2,t_j}){1\over{\sqrt 2}}(b^{\dagger}_{1,t_i}+b^{\dagger}_{2,t_i})|0>\eeq

Hence, out of the $n^2$ orthogonal, equal magnitude terms in the original state $ a^{\dagger}_2
a^{\dagger}_1|0>={1\over{\sqrt n}}\sum_{j=1}^n a^{\dagger}_{2, t_j} {1\over{\sqrt n}} \sum _{i=1}^n
a^{\dagger}_{1,t_i}|0>$ only $n$ equal magnitude, orthogonal terms (the terms in which $i=j$) lead to
scattering, while all others behave as in the photon case. Therefore, in the case of long (relative to the
scattering lengths) wave packets, atoms behave identically to photons, up to corrections of $1/n$.  That is,
instead of (\ref{atomsmandeldip}), two atoms in long wave packets, impinging on the two sides of a beamsplitter
actually lead to an Hong-Ou-Mandel dip, \beq a^{\dagger}_2 a^{\dagger}_1|0>\rightarrow {1\over
2}\bigl((b^{\dagger}_1)^2-(b^{\dagger}_2)^2\bigr)|0>+O(1/n)\eeq where $O(1/n)$ denotes corrections to the
wavefunction with norm  of the order $1/n$.

A similar analysis can be done for the case of more atoms impinging on a beam-splitter. As long as the
wave-packets are long relative to the scattering length, scattering is very unlikely to occur, and atoms behave
identically to photons. Hence, in this regime, throughout the whole KLM computer, atoms behave like photons, up
to corrections that can be made as small as we want by enlarging the size of the wave-packets. Furthermore, note
that by construction the original photonic KLM computation already had to accept errors (the gates are
probabilistic); the errors introduced by scattering in the case of atom KLM computation can be dealt with in
similar ways to the original errors.

In practice, the cross-section for neutral atom scattering is of the order of angstroms ($10^{-10}$m), so it is
conceivable that atoms can be prepared in long enough wave-packets to make the total amount of errors small
enough.

In conclusion, in this note I argued that KLM quantum computation can be performed not only with photons, as
originally envisaged, but also with bosonic neutral atoms or bosonic ions. Versus other quantum computation
schemes with neutral atoms or ions, KLM computation has the advantage that no controlled interaction between
their internal levels are required - only control over the center of mass movement is needed. Versus photonic
KLM computation, this scheme has the advantage that single atom/ion sources are more natural than single photon
sources, and the detectors are far more efficient than single photon detectors. At the same time, the neutral
atom/ion ``optics'' required for this scheme is far less developed that standard photonic optics, so it is not
at all clear which scheme is more advantageous in the long run. Ultimately, all kinds of details of technology
will have the decisive role.

\begin{acknowledgments}
The author would like to thank A. Short for very useful discussions. This work was supported by the UK EPSRC
grant GR/527405/01 and the UK EPSRC ÒQIP IRCÓ project.\end{acknowledgments}

\end{document}